\documentclass[11pt,a4paper]{article}

\usepackage[margin=2.4cm]{geometry}
\usepackage[T1]{fontenc}
\usepackage{lmodern}
\usepackage{microtype}
\usepackage{amsmath,amssymb}
\usepackage{booktabs}
\usepackage{multirow}
\usepackage{graphicx}
\usepackage{xcolor}
\usepackage{caption}
\usepackage{subcaption}
\usepackage{listings}
\usepackage{tikz}
\usepackage{pgfplots}
\usepackage{pgfplotstable}
\pgfplotsset{compat=1.18}
\usepgfplotslibrary{fillbetween}
\usetikzlibrary{arrows.meta,positioning,fit,calc}
\usepackage{authblk}
\usepackage[numbers,sort&compress]{natbib}
\usepackage[hidelinks,colorlinks=true,linkcolor=blue!50!black,citecolor=blue!50!black,urlcolor=blue!50!black]{hyperref}
\usepackage{xurl}
\usepackage{orcidlink}

\definecolor{cwake}{RGB}{0,110,150}
\definecolor{cfoundry}{RGB}{200,90,20}
\definecolor{cechidna}{RGB}{90,90,90}
\definecolor{cagent}{RGB}{170,0,140}

\newcommand{\wake}{Wake}
\newcommand{\revm}{\texttt{revm}}
\newcommand{\pp}{\,pp}
\newcommand{\tablenote}[1]{\par\smallskip\begin{minipage}{\linewidth}\footnotesize\raggedright #1\end{minipage}}

\title{\wake{}~5: A Python Framework for Solidity Testing, Fuzzing, and Static Analysis}

\author[1,2]{Josef Gattermayer\,\orcidlink{0009-0008-0293-4209}}
\author[1]{Michal P\v{r}evr\'{a}til}
\affil[1]{ack3 (ACK3 LTD), ack3.ai}
\affil[2]{Faculty of Information Technology, Czech Technical University in Prague}
\date{November 2026}

\begin{document}
\maketitle

\begin{abstract}
Smart-contract fuzzers differ by an order of magnitude in throughput and by design in how they construct inputs, yet published comparisons rarely separate the two. We present \wake{}~5, an open-source Python framework for Solidity testing, fuzzing and static analysis, and use it to locate the binding constraint of stateful smart-contract fuzzing. \wake{}~5 replaces the JSON-RPC connection to an external node with an in-process EVM (\revm{}) and exposes its static analysis to language-model agents through a Model Context Protocol (MCP) server. On the Daedaluzz maze benchmark (153 reachable assertions, guard constants below $2^{32}$), \wake{}~5 executes 29{,}931 transactions per second, on par with Foundry (30{,}186) and 11.5$\times$ faster than Echidna (2{,}600). In eight-hour campaigns, however, the three tools plateau at medians of 75, 71 and 70 assertions, and their union reaches 93. We show that \wake{}'s integer generator places 51\% of its probability mass below $2^{32}$ and that, once shallow assertions are exhausted, the generator prior rather than throughput bounds every tool. A harness written by a language-model agent, which derives inputs with a constraint solver, reached all 153 assertions in each of five campaigns executed by \wake{}, in 33 to 79 minutes. A complementary ablation in an agentic audit of a production Go codebase found that static-analysis tools improved recall for none of three models, raised precision for all three, and increased language-model cost by up to 12.7\%. We discuss correlated oracles and non-deterministic completion, and argue that the fuzzer's role is shifting from search to execution.
\end{abstract}


\section{Introduction}
\label{sec:intro}

Smart contracts on the Ethereum Virtual Machine (EVM)~\cite{wood2014ethereum} are immutable once deployed and directly custody value, which makes pre-deployment testing unusually consequential. A companion dataset of 135 decentralised-finance (DeFi) incidents from the first half of 2026 found that, among the 68 incidents with identified pre-incident audit history, 94\% of reported losses came from attack paths outside every identified pre-incident audit scope~\cite{gattermayer2026incidents}. Fuzzing is the dominant dynamic technique for finding such paths before an attacker does: Echidna~\cite{grieco2020echidna}, Foundry~\cite{foundry}, Medusa~\cite{medusa}, Harvey~\cite{wustholz2020harvey} and ItyFuzz~\cite{shou2023ityfuzz} all generate transaction sequences and check assertions or invariants, and Daedaluzz~\cite{daedaluzz2023,wustholz2023benchmarking} was introduced to compare them on a common footing.

Two questions about these tools remain poorly separated in the literature. The first is architectural: does the execution engine matter? Python-based frameworks such as Brownie~\cite{brownie} and Ape~\cite{ape} drive an external node over JSON-RPC, while Foundry and Echidna execute in-process; a study of eleven contract fuzzers found throughput and seed quality pivotal and a standalone EVM the main throughput lever, but left property-based tools such as Echidna and Foundry outside its scope~\cite{wu2024arewethereyet}. The second is methodological: once a fuzzer is fast, what limits the bugs it finds? Coverage-guided greybox fuzzing~\cite{bohme2016aflfast} answers this for C programs, but stateful contracts with arithmetic guards over multi-transaction state behave differently, and a third participant has entered the workflow: language-model agents that write the harness~\cite{zhang2024fuzzdriver,xia2024fuzz4all}.

This paper describes \wake{}~5 and uses it to measure both questions. \wake{} is an open-source framework in which tests, fuzz harnesses and static analyses are written in Python against types generated from the Solidity compiler's output~\cite{wake2026repo}. The framework has been used in commercial security reviews since 2023 and its fuzz tests are credited with critical and high-severity findings in several public audit reports~\cite{ipor2023audit,pwn2023audit,lido2025csm}. \wake{}~5 is the first release whose default execution engine is an EVM embedded in the Python process rather than an external node reached over a socket.

\paragraph{Contributions.}
\begin{enumerate}
  \item We describe the design of \wake{}~5: a version-normalising intermediate representation over the Solidity abstract syntax tree (AST), an in-process EVM (\revm{}~\cite{revm}) bound to Python through PyO3~\cite{pyo3}, a bounded block-history window that makes multi-hour campaigns memory-stable, \emph{manually guided fuzzing} (MGF) as the harness model, a two-phase stateful shrinker, and a Model Context Protocol (MCP) server~\cite{mcp} that exposes static analysis to agents (Section~\ref{sec:design}).
  \item We measure \wake{}~5, Echidna~2.3.3 and Foundry~1.7.1 on an L32 maze set generated with Daedaluzz (guard constants below $2^{32}$): 75 eight-hour campaigns plus 45 twenty-minute throughput runs, one physical core each. \wake{}~5 matches Foundry's throughput, all three tools end within five assertions of one another, and an exact analysis of \wake{}'s integer generator shows why the prior rather than the engine sets the hit rate on constant-guarded code (Sections~\ref{sec:method}--\ref{sec:prior}).
  \item We report five agent-written harness campaigns that complete the benchmark in 33 to 79 minutes by moving guard satisfaction into a satisfiability modulo theories (SMT) solver and navigation into breadth-first search, and we analyse what changed and what did not (Section~\ref{sec:agent}).
  \item We report an ablation of static-analysis tools in an agentic security review of a production Go codebase (three models, 134 paired cases: recall unchanged, precision up, cost up) and identify three failure modes of agent-written harnesses observed in practice: correlated oracles, non-deterministic completion and priority inversion (Sections~\ref{sec:ablation} and~\ref{sec:discussion}).
\end{enumerate}

\section{Background and Related Work}
\label{sec:background}

\paragraph{Smart-contract fuzzers.}
Echidna~\cite{grieco2020echidna} is a Haskell property-based fuzzer that derives an input grammar from the contract's application binary interface (ABI), maintains a coverage-guided corpus and mines constants from the bytecode into a value dictionary. Foundry's \texttt{forge}~\cite{foundry} executes Solidity-written fuzz and invariant tests on \revm{} with typed random inputs and value dictionaries. Medusa~\cite{medusa} is a parallel, coverage-guided mutational fuzzer with Solidity properties. Harvey~\cite{wustholz2020harvey} adds input prediction to greybox search; ItyFuzz~\cite{shou2023ityfuzz} snapshots EVM state to reach deep states; Smartian~\cite{choi2021smartian} steers generation with static and dynamic data-flow information; sFuzz~\cite{nguyen2020sfuzz} combines AFL-style fuzzing with a multi-objective adaptive strategy for hard-to-cover branches; ContractFuzzer~\cite{jiang2018contractfuzzer} targets known vulnerability oracles. \wake{} differs in where the harness lives: flows, invariants and the differential model (a Python re-implementation of the expected behaviour that serves as the test oracle) are Python code, and the engine is a dependency of the test process rather than the other way round.

\paragraph{Benchmarks.}
Fuzzle~\cite{lee2022fuzzle} generates C programs from maze puzzles so that reachability requires satisfying chains of constraints. Daedaluzz~\cite{daedaluzz2023,wustholz2023benchmarking} applies the idea to Solidity: a contract keeps a position on a grid across transactions, moves are guarded by arithmetic conditions over eight \texttt{uint64} parameters, and the score is the number of distinct assertion violations; its runners cover Echidna, Foundry, Harvey, Hybrid-Echidna and ItyFuzz. Olympia~\cite{chadt2024olympia} is a second Solidity maze generator on which ItyFuzz, Echidna, Medusa and Foundry were compared at one-hour timeouts, ItyFuzz solving the most mazes. Klees et~al.~\cite{klees2018evaluating} and Schloegel et~al.~\cite{schloegel2024sok} set the methodological bar we follow: multiple seeds, long campaigns, per-time series rather than endpoints, and statistical tests when comparing. Evaluations of language-model agents on contract security have recently appeared~\cite{wang2026evmbench,gattermayer2026frontier}, as have large-scale evaluations of language-model-written fuzz drivers for C libraries~\cite{zhang2024fuzzdriver}; none that we are aware of puts mainstream EVM fuzzers and an agent-written harness on one benchmark with the harness executed by one of the engines.

\paragraph{Property-based testing and shrinking.}
\wake{}'s flow model descends from QuickCheck~\cite{claessen2000quickcheck} and Hypothesis~\cite{maciver2019hypothesis}: generators produce inputs, properties are checked after every step, and failing sequences are reduced. Reduction of stateful sequences is a delta-debugging problem~\cite{zeller2002simplifying,regehr2012creduce} with the added constraint that the Python-side model must be replayed alongside the chain.

\paragraph{Static analysis.}
Slither~\cite{feist2019slither} compiles Solidity to an SSA-form intermediate representation and runs pattern detectors. \wake{} keeps the compiler's AST shape and adds cross-version normalisation, control-flow and data-dependency graphs, and a detector and printer API; the same representation serves a language server~\cite{lsp} and, since \wake{}~5, an MCP server~\cite{mcp}.

\paragraph{Agents that write fuzzers.}
Fuzz4All~\cite{xia2024fuzz4all} and TitanFuzz~\cite{deng2023titanfuzz} use language models to generate inputs or harness code for compilers and deep-learning libraries, and Zhang et~al.~\cite{zhang2024fuzzdriver} measure language-model-written fuzz drivers at scale. For contracts, LLM4Fuzz~\cite{shou2024llm4fuzz} has a language model prioritise what a fuzzer explores, while PropertyGPT~\cite{liu2025propertygpt} and SmartInv~\cite{wang2024smartinv} have it write properties and invariants that a prover or fuzzer then checks, which is the oracle question Section~\ref{sec:discussion} returns to. Echidna's agents preview exposes a running campaign to language models over MCP~\cite{echidna230}, and slither-mcp~\cite{slithermcp} exposes Slither's analyses; \wake{}'s server exposes static facts only (Section~\ref{sec:static}). The harness in Section~\ref{sec:agent} is closer to a program-synthesis result: the agent produced a solver-backed navigator rather than random inputs, and the fuzzer's role reduced to executing and replaying candidates.

\section{The Design of \wake{}~5}
\label{sec:design}

\begin{figure}[t]
\centering
\resizebox{\textwidth}{!}{%
\begin{tikzpicture}[
  box/.style={draw,rounded corners=2pt,align=center,minimum height=9mm,minimum width=34mm,inner sep=4pt,font=\small},
  py/.style={box,fill=cwake!8},
  rs/.style={box,fill=cfoundry!10},
  ext/.style={box,fill=gray!10},
  arr/.style={-{Latex[length=2mm]},thick}
]
\node[ext] (solc) at (0,0) {solc\\(one process per version)};
\node[py] (ir) at (4.6,0) {IR (\texttt{wake.ir})\\version-normalised AST};
\node[py] (lsp) at (10.2,1.5) {Language server\\(\texttt{wake lsp})};
\node[py] (det) at (10.2,0) {Detectors and printers\\control-flow and data-dependency graphs};
\node[py] (mcp) at (10.2,-1.5) {MCP server\\(\texttt{wake mcp})};
\node[py] (pyt) at (4.6,-3.3) {pytypes\\generated Python classes};
\node[py] (test) at (10.2,-3.3) {Tests and fuzz harnesses\\pytest, \texttt{FuzzTest}, flows, invariants};
\node[rs] (wrs) at (10.2,-5.1) {\texttt{wake\_rs} (Rust, PyO3)\\accounts, \texttt{Address}, ABI, RNG, snapshots};
\node[rs] (revm) at (15.4,-5.1) {\revm{}\\in-process EVM};
\node[ext] (rpc) at (4.6,-5.1) {Anvil / Hardhat\\over JSON-RPC (optional)};
\draw[arr] (solc) -- (ir);
\draw[arr] (ir.east) -- (lsp.west);
\draw[arr] (ir) -- (det);
\draw[arr] (ir.east) -- (mcp.west);
\draw[arr] (ir) -- (pyt);
\draw[arr] (pyt) -- (test);
\draw[arr] (test) -- (wrs);
\draw[arr] (wrs) -- (revm);
\draw[arr,dashed] (test.south west) -- (rpc.north east);
\end{tikzpicture}}
\caption{\wake{}~5 architecture. Blue boxes denote Python components; orange boxes denote Rust components. Transactions use the in-process \texttt{wake\_rs}--\revm{} path by default, or the JSON-RPC path to Anvil or Hardhat.}
\label{fig:arch}
\end{figure}
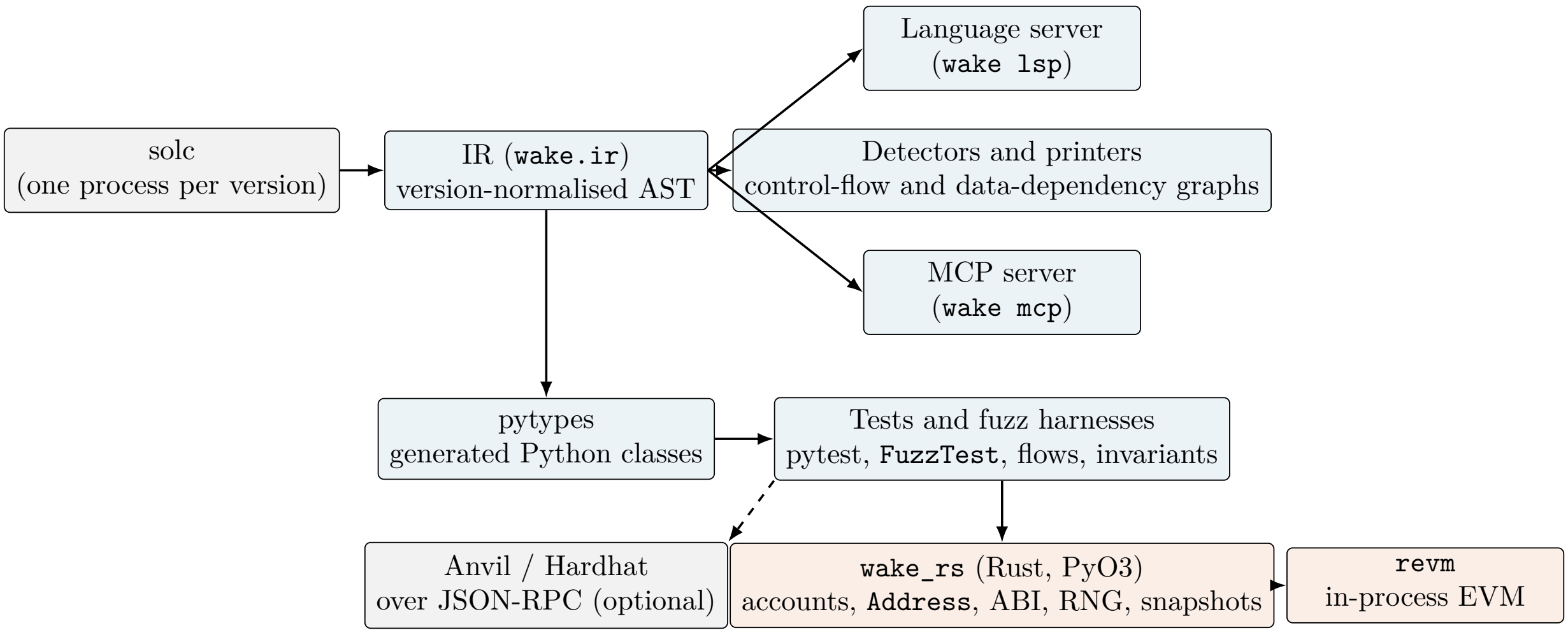

\wake{} is organised around one intermediate representation (IR) built from the AST emitted by the Solidity compiler (\texttt{solc}), and several consumers of it (Figure~\ref{fig:arch}). The testing framework generates Python types (\emph{pytypes}) from the IR and executes tests against an EVM; the static-analysis framework, the language server and the MCP server query the IR directly.

\subsection{Intermediate representation}
\label{sec:ir}

Solidity couples source to compiler version: a file declaring \texttt{pragma solidity 0.8.4} must be compiled with \texttt{solc}~0.8.4, ASTs differ across versions in node attributes and known defects~\cite{solidity}, and one project routinely spans several versions with the same library file in several compilation units. \wake{}'s IR keeps the AST's node types and structure but merges all compilation units into one model, resolves references across units, corrects known version-specific defects (missing referenced declarations of enum values before 0.8.2, mislabelled hexadecimal string literals before 0.7.0), and records the compiler version ranges of each source unit. A file compiled in several units gets one canonical AST into which the other units index their nodes. Analyses therefore traverse one tree per source file without version-specific code. The IR is pickled to disk and recompiled incrementally, which is what lets the language server answer within the editing loop.

\subsection{Hybrid Python/Rust execution}
\label{sec:hybrid}

A testing framework has three layers: the test specification, the framework core, and the EVM. Foundry writes all three in Solidity and Rust; Hardhat~\cite{hardhat} uses TypeScript for the first two; Brownie and Ape use Python for the first two and reach an external node for the third. Solidity as a test language lacks floating-point arithmetic and numerical libraries, which matters for differential models of financial protocols, and it invites circular dependency between implementation and test. Python as a test language has those libraries and is represented in StarCoder's filtered training corpus at roughly seventy times the volume of Solidity~\cite{li2023starcoder}. What Python-based frameworks lacked was throughput: every transaction crossed a socket.

\wake{}~5 embeds \revm{}~\cite{revm} through a Rust extension module (\texttt{wake\_rs}) built with PyO3~\cite{pyo3}, the Rust binding layer for the Python interpreter, and the Alloy Ethereum libraries~\cite{alloy}. Transactions execute synchronously in the test process; there is no node, no RPC endpoint and no serialisation of calldata over a socket. The module also implements the \texttt{Address} type, account management including Trezor hardware-wallet signing, ABI encoding, an internal seeded random-number generator (RNG) for chain-level values such as block hashes, and chain state snapshots. Fuzz-test values are still drawn from a seeded Python generator; both generators are captured by snapshots so that shrinking replays deterministically (Section~\ref{sec:shrink}). The JSON-RPC path is retained for Anvil and Hardhat; Ganache support was dropped.

Two engineering decisions were necessary for eight-hour campaigns. First, \wake{}~4 retained every block and transaction for the whole run; under \revm{} a twenty-minute campaign retained roughly 1.6~GB per million transactions and exhausted the container's memory. \wake{}~5 keeps only the newest \texttt{block\_history} blocks (default 256, with a small hysteresis margin before pruning) and their transactions. Access outside the window raises \texttt{HistoryPrunedError}; transaction objects that the test keeps a reference to remain valid, while call traces and console logs, which re-execute the transaction, require the block to be retained. Second, reverted-error objects held reference cycles through the chain; breaking them was required for the memory profile to stay flat.

\subsection{Manually guided fuzzing}
\label{sec:mgf}

\wake{}'s harness model is \emph{manually guided fuzzing} (MGF). A harness (a \emph{fuzz test} in \wake{}'s vocabulary) is a Python class deriving from \texttt{FuzzTest}. Its methods decorated with \texttt{@flow} are atomic steps, each usually one transaction, chosen at random according to weights and preconditions; methods decorated with \texttt{@invariant} are checked after every flow or every $k$ flows. The invariants and the differential model are the harness's test oracles, the parts that decide whether a sequence passes; we write \emph{oracle} when either is meant. A \emph{sequence} is one independent run of a fixed number of flows from a fresh deployment; chains are reset between sequences, and \texttt{pre\_sequence} deploys and initialises the target before each one. Flow arguments are generated from Python type hints (\texttt{uint64}, \texttt{Address}, pytypes structs and enums), and the harness keeps a differential model of the contract in Python state and compares it with on-chain state in its invariants. Listing~\ref{lst:mgf} shows the shape on an illustrative bank contract. The human supplies the transaction vocabulary, value strategy, differential model and assertions; \wake{} randomises within those choices.

\begin{lstlisting}[language=Python,caption={Manually guided fuzz test for a bank contract.},label={lst:mgf}]
class BankFuzz(FuzzTest):
    def pre_sequence(self):
        self.user = random_account()
        self.bank = Bank.deploy()
        self.expected = 0

    @flow(weight=3)
    def deposit(self):
        amount = random_int(1, 10**18)
        self.bank.deposit(from_=self.user, value=amount)
        self.expected += amount

    @flow(precondition=lambda s: s.expected > 0)
    def withdraw(self):
        amount = random_int(1, self.expected)
        self.bank.withdraw(amount, from_=self.user)
        self.expected -= amount

    @invariant()
    def balance_matches(self):
        assert self.bank.balanceOf(self.user) == self.expected
\end{lstlisting}

Three features distinguish MGF from black-box or property-only fuzzing. (i)~\emph{Revert expectations are explicit.} A flow wraps a call in a \texttt{may\_revert} or \texttt{must\_revert} context manager that names the errors it tolerates; any other revert (a transaction that aborts and rolls back its effects) fails the test. Property-based fuzzers typically treat a reverted transaction as an uninteresting input, whereas an unexpected revert is how denial-of-service and stuck-state bugs surface. (ii)~\emph{The oracle is a model, not only an invariant.} Differential models in Python have found rounding and accounting errors that no single-transaction invariant expresses~\cite{ipor2023audit}. (iii)~\emph{The method is iterative.} A harness grows over many develop--run--debug cycles; bugs are typically found while the harness is being built, before any long campaign runs. Section~\ref{sec:agent} treats the agent as a participant in exactly this loop.

Flow arguments of integer type are drawn from a prior that is not uniform. With probability 0.05 the generator returns one of five edge values $\{0,\,1,\,2^{n}-1,\,2^{n-1},\,2^{n-1}-1\}$ for an unsigned $n$-bit type (signed types add two negative edges); otherwise it draws a bit length $b$ uniformly from $\{1,\dots,n\}$ and then a uniform value in $[0,2^{b})$, that is, a value of at most $b$ bits. Section~\ref{sec:prior} quantifies the consequences.

\subsection{Stateful shrinking}
\label{sec:shrink}

When a sequence fails, \wake{} shrinks it, that is, it searches for a shorter sequence that still fails, in two phases. Phase~1 groups flows by kind, orders kinds by call count, and attempts to remove every call of a kind at once, re-running the sequence after each attempt. Phase~2 walks the remaining sequence from the start and removes flows one at a time, restoring a flow whenever the failure disappears. Both phases must keep the Python-side model consistent with the chain, so \wake{} snapshots the chain state and deep-copies the harness object at checkpoints, and restarts replay from the nearest checkpoint rather than from the beginning. A shortcut rule replaces the failing suffix with an earlier occurrence of the same error when one is found during replay. In \wake{}~5 the shrinker was reworked for the \revm{} engine: each checkpoint captures the Python random state and the engine's internal RNG state together with the chain snapshot, so that replay reproduces the same generated values, and a performance gap between chain snapshots and Python state copies was closed. Command-line switches select whether the reduced sequence must fail in the same flow with the same exception or merely fail. Unlike Echidna, \wake{} does not simplify argument values; unlike Foundry, it does not stop at transaction removal only.

\subsection{Static analysis, language server and MCP server}
\label{sec:static}

The static-analysis framework provides a visitor API over the IR, per-function control-flow graphs, data-dependency information, and 26 built-in detectors and 19 printers (5.0.0rc3). Detectors range from local patterns (unchecked return values, \texttt{tx.origin} authentication, unsafe \texttt{delegatecall}) through known compiler defects to graph-based queries such as reentrancy, detected as state written on a control-flow path after an external call. Findings can be suppressed inline, exported as JSON and filtered against a database of known deployed contracts, and the language server surfaces them, together with navigation, hover and rename, through the Language Server Protocol~\cite{lsp}.

\wake{}~5 adds an MCP server (\texttt{wake mcp}) that exposes the same static facts to language-model agents~\cite{mcp}: contracts and functions, definitions and references, C3 linearisation (the method-resolution order across inheritance), storage layout, state changes performed by a function, lookups by four-byte function selector, and external-call sites. Every tool answers from the compiled IR as compact plain text; on one representative response the plain-text form used 83 tokens against 199 for JSON (\texttt{o200k\_base} tokenizer). Section~\ref{sec:ablation} measures whether tools of this kind change audit outcomes.

\section{Evaluation Methodology}
\label{sec:method}

We ask five questions.
\textbf{RQ1} (throughput): how many transactions per second does \wake{}~5 execute compared with Echidna and Foundry on the same contracts and hardware?
\textbf{RQ2} (coverage): how many distinct reachable assertions does each tool violate in eight hours, and how do the curves evolve?
\textbf{RQ3} (prior): how much of the input space does \wake{}'s integer prior concentrate on, and what does that imply for constant-guarded assertions?
\textbf{RQ4} (agent): what does an agent-written harness achieve on the same benchmark, and where did the work move?
\textbf{RQ5} (static-analysis tools): do static-analysis tools that return deterministic code-structure facts change recall, precision or cost in an agentic security review?

\subsection{Benchmark}
\label{sec:benchmark}

Daedaluzz~\cite{daedaluzz2023} generates Solidity contracts that hold a position $(x,y)$ on a $7{\times}7$ grid across transactions. Four external functions (\texttt{moveNorth}, \texttt{moveSouth}, \texttt{moveEast}, \texttt{moveWest}) each take eight \texttt{uint64} parameters $p_0,\dots,p_7$, move the position, and evaluate a cell-specific chain of arithmetic guards; some cells contain an \texttt{assert(false)} reachable only when the guards hold, some cells are walls, and some guards are unsatisfiable. Arithmetic is \texttt{unchecked}, so products and sums wrap modulo $2^{64}$. The score of a campaign is the number of distinct assertions violated. Listing~\ref{lst:cell} shows one cell from maze~0. Reaching its assertion requires five transactions (four transit moves and the triggering move), with inputs satisfying an inequality and an exact relation between two 64-bit values.

\begin{lstlisting}[caption={Guard conditions and assertion at cell $(0,5)$ of L32 maze~0.},label={lst:cell}]
if (x == 0 && y == 5) {
  if (p6 != uint64(uint64(2398633357) + p1)) {
    if (p4 == uint64(p5 * p5)) {
      emit AssertionFailed("5"); assert(false);
    }
  }
  return 5;
}
\end{lstlisting}

The published Daedaluzz mazes draw guard constants from $[0,64]$, which favours any generator biased towards small values. We therefore generated an \emph{L32} set: five mazes from fixed seeds 0--4 with the generator's constant bound raised to $2^{32}-1$ while inputs remain \texttt{uint64}. Solving every guard chain with Z3~\cite{demoura2008z3} (each parameter a 64-bit bit-vector, wrap-around preserved) and replaying each satisfying model against a fresh deployment established that exactly 153 assertions are reachable: $36+24+32+27+34$ across the five mazes. That count is the denominator throughout. All contracts were compiled with \texttt{solc}~0.8.19, optimizer enabled with 200 runs.

\subsection{Tools and harnesses}
\label{sec:tools}

Table~\ref{tab:tools} lists the tools. Each tool used its own idiomatic harness: Echidna in exploration mode on the stock contract, an assertion counting as violated when its \texttt{emit AssertionFailed} line is first covered (the event itself is discarded by the revert), which is how the benchmark's own runner scores it; Foundry with the generator's \texttt{.foundry.sol} invariant variant, one invariant function per assertion; \wake{} with a \texttt{FuzzTest} of four flows, one per move function, each taking eight \texttt{uint64} flow arguments and wrapped in \texttt{may\_revert()}, and an invariant with period 100 that polls the contract's \texttt{found} mapping and prints the first-seen time of every newly violated assertion. The \wake{} harness is deliberately unguided: no value strategy, no model, no preconditions. It measures the engine and the default generator, not the MGF method.
Neither Echidna's symbolic-execution mode, off by default since 2.3.0~\cite{echidna230}, nor Foundry's corpus-guided invariant mode, opt-in since 1.3.0~\cite{foundry130}, was enabled; Foundry ran with the benchmark's dictionary settings (storage and push-byte dictionaries, invariant depth 100, 500 runs per invocation, reverts tolerated). Both tools therefore ran as random generators with value dictionaries, and the comparison is between random-input fuzzers, not between solver-assisted ones.

\begin{table}[t]
\centering\footnotesize
\caption{Tool versions and benchmark configurations}
\label{tab:tools}
\begin{tabular}{@{}llp{7.6cm}@{}}
\toprule
Tool & Version & Notes \\
\midrule
Echidna & 2.3.3 & exploration mode, coverage-guided corpus, sequence length 100; patched to timestamp the first coverage of each assertion line \\
Foundry (\texttt{forge}) & 1.7.1 (commit \texttt{4072e487}) & invariant tests, one per assertion; \texttt{solc} pre-fetched into the image \\
\wake{} & 5.0.0rc2 (commit \texttt{d33810e}) & \revm{} engine; first build with bounded block history \\
Agent-written harness & five campaigns & Claude Opus~4.8 (\texttt{xhigh}); GPT-5.5 (\texttt{high}) researcher configured but unavailable in campaign~5; Python harness executed by \wake{} \\
\bottomrule
\end{tabular}
\end{table}

\subsection{Campaigns and measurement}
\label{sec:campaigns}

\emph{Coverage campaigns} (RQ2): 5 seeds $\times$ 5 mazes $\times$ 3 tools $=75$ campaigns of eight hours. Each campaign is a Docker container pinned to one physical core of a 16-core AMD Ryzen~AI~MAX+~395 workstation, with simultaneous multithreading siblings left idle; seeds 0--2 ran in August 2026 and seeds 3--4 in October 2026 from the same pinned images. A three-seed cut of these campaigns (medians 73, 72 and 71; union 90) was reported in a research note~\cite{prevratil2026fuzzers}; seeds 3 and 4 were added for this paper and every number here is recomputed over five seeds. Per-run series record the cumulative number of distinct violated assertions every ten seconds. The reported endpoint of a tool is the median across seeds of the per-seed sum over the five mazes.

\emph{Throughput runs} (RQ1): a separate set of 3 seeds $\times$ 5 mazes per tool with a twenty-minute target, 15 runs per tool, one core per run. Transaction counts come from each tool's own reporting: \wake{} counts flows in the harness, Echidna prints its total call count, and Foundry's per-invariant call counts are summed over invocations. Active time excludes compilation and setup. We report medians over the 15 runs. Throughput is machine-specific; the ratios, not the absolute rates, are the result.

\emph{Agent campaigns} (RQ4): five end-to-end runs of the agent on the same five mazes, each one process covering harness construction, solving and execution. For the August 2026 campaign the milestones are the wall-clock times at which the harness's replayed reproductions reached 150 and 153 distinct assertions; for the three campaigns of 5~October 2026 they are the first-hit timestamps of the recorded findings, one of which also records from-scratch reproduction. For the campaign of 26~October 2026, run under a fixed protocol after the first draft of this paper (byte-identical L32 fixtures, \wake{} at the measured commit, finding records, harness and per-conversation token accounting retained), they are the earliest first-visit timestamp per assertion across its fuzz passes. Time is measured from the agent's start, so it includes harness construction and solver work. Every campaign used Claude Opus~4.8 at \texttt{xhigh}, with GPT-5.5 at \texttt{high} configured as a research sub-agent in the same runner. In the fifth campaign, both retained research-sub-agent calls failed with a model-availability error; the main agent completed the harness without a successful researcher response. The agent has no stopping rule tied to the benchmark: three campaigns ended on their own after further work on a target with nothing left to find, 59, 123 and 170 minutes after their last assertion, and two were stopped by the operator (Table~\ref{tab:agentcost}). Wall time and language-model cost are therefore reported per campaign as end-to-end totals that include that overrun; the completion minute is the comparable quantity. The finding records from 5~October were inspected at the time and not retained: the agent's working directory is cleared between campaigns so that no run can reuse an earlier run's findings, and the session step logs are the surviving record. Five runs show the spread of completion times without estimating it.

\emph{Statistics}: differences between tools in the per-seed eight-hour endpoints are tested with the exact two-sided Mann--Whitney $U$ test and reported with the Vargha--Delaney $\hat{A}_{12}$ effect size~\cite{vargha2000a12}, following the guidelines of Arcuri and Briand~\cite{arcuri2014hitchhiker}; with five seeds per tool the smallest attainable $p$ is 0.008. Paired comparisons of binary outcomes in the ablation (RQ5, Section~\ref{sec:ablation}) use exact McNemar tests~\cite{mcnemar1947}. For the paired change $d=(b-c)/n$, we report normal-approximation 95\% intervals $d\pm1.96\sqrt{[(b+c)/n-d^2]/n}$, where $b$ and $c$ count cases correct only with tools on and only with tools off, respectively; $n$ is 55 for recall and 79 for correct rejection.

\section{Results}
\label{sec:results}

\subsection{RQ1: Throughput}
\label{sec:throughput}

\begin{figure}[t]
\centering
\begin{tikzpicture}
\begin{axis}[
  width=0.48\textwidth, height=5.2cm,
  ybar, bar width=14pt,
  symbolic x coords={Foundry, Wake 5, Echidna}, xtick=data,
  ymin=0, ymax=36000,
  ylabel={transactions / s}, ylabel near ticks,
  nodes near coords, nodes near coords style={font=\scriptsize},
  scaled y ticks=false, yticklabel style={/pgf/number format/fixed,/pgf/number format/1000 sep={,}},
  enlarge x limits=0.3, tick label style={font=\small}, label style={font=\small},
  title={\small (a) Throughput}
]
\addplot[fill=cwake!30,draw=cwake] coordinates {(Foundry,30186) (Wake 5,29931) (Echidna,2600)};
\end{axis}
\end{tikzpicture}
\hfill
\begin{tikzpicture}
\begin{axis}[
  width=0.48\textwidth, height=5.2cm,
  ybar, bar width=14pt,
  symbolic x coords={Foundry, Wake 5, Echidna}, xtick=data,
  ymin=0, ymax=4,
  ylabel={assertions / $10^6$ tx}, ylabel near ticks,
  nodes near coords, nodes near coords style={font=\scriptsize},
  enlarge x limits=0.3, tick label style={font=\small}, label style={font=\small},
  title={\small (b) Assertions per million transactions}
]
\addplot[fill=cfoundry!30,draw=cfoundry] coordinates {(Foundry,0.32) (Wake 5,0.37) (Echidna,3.36)};
\end{axis}
\end{tikzpicture}
\caption{Transaction throughput and assertion discovery on the L32 mazes. Panels show medians over 15 runs per tool with a twenty-minute target: (a) transactions per second; (b) distinct assertions violated per million transactions.}
\label{fig:throughput}
\end{figure}
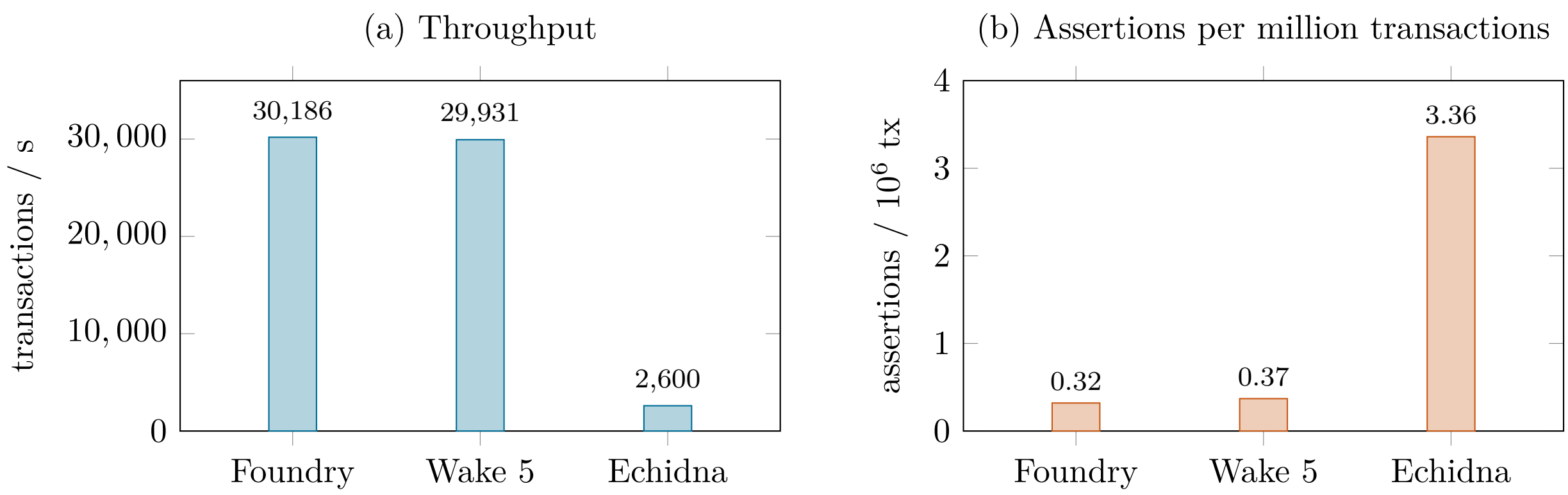

Table~\ref{tab:throughput} and Figure~\ref{fig:throughput} give the twenty-minute throughput runs. \wake{}~5 executed a median 29{,}931 transactions per second, Foundry 30{,}186 and Echidna 2{,}600. Two tools that embed \revm{} run within one percent of each other; the Haskell-hosted Echidna is 11.5--11.6$\times$ slower. This is the architectural result: moving the EVM into the test process removes the gap between Python-hosted and Rust-hosted fuzzing.

\begin{table}[t]
\centering\small
\caption{Transaction throughput and assertion discovery on L32 mazes}
\label{tab:throughput}
\begin{tabular}{@{}lrrrrr@{}}
\toprule
Tool & Transactions & Tx/s & vs.\ Echidna & Distinct assertions & Assertions / $10^6$ tx \\
\midrule
Foundry & 36{,}655{,}640 & 30{,}186 & 11.6$\times$ & 12 & 0.32 \\
\wake{}~5 & 35{,}895{,}579 & 29{,}931 & 11.5$\times$ & 13 & 0.37 \\
Echidna & 3{,}119{,}791 & 2{,}600 & 1.0$\times$ & 11 & 3.36 \\
\bottomrule
\end{tabular}
\tablenote{Medians over 15 runs per tool, with a twenty-minute target and one core per run. Asserts counts distinct assertions violated; Asserts/$10^6$\,tx uses these short runs, not the eight-hour campaigns.}
\end{table}

Per transaction the ranking inverts. Echidna violated 3.36 distinct assertions per million transactions against 0.37 for \wake{} and 0.32 for Foundry. Coverage guidance makes each transaction more informative; the \revm{}-based tools compensate with volume. In twenty minutes the outcome is nearly the same: the median run violated 11, 13 and 12 assertions on its maze.

\subsection{RQ2: Eight-hour coverage}
\label{sec:coverage}

Figure~\ref{fig:coverage} shows the median curves and Table~\ref{tab:endpoints} the per-seed endpoints. After eight hours the medians are \wake{}~75, Foundry~71 and Echidna~70 of 153 reachable assertions; the fifteen per-seed sums range from 67 to 75. The ordering is consistent across seeds: \wake{} exceeds Foundry in four of five seeds and Echidna in all five (Mann--Whitney $p=0.040$, $\hat{A}_{12}=0.90$, and $p=0.016$, $\hat{A}_{12}=0.98$), whereas Foundry and Echidna are not separated ($p=0.25$, $\hat{A}_{12}=0.76$). The effect is nonetheless small in absolute terms, four assertions of 153. Most of each total arrives within the first hour (\wake{}~68, Foundry~61, Echidna~61 at one hour), after which the curves add a tail of seven to ten assertions over seven hours. The findings overlap heavily: the union of assertions found by any tool in any campaign is 93, leaving 60 that no campaign of the three tools reached.

\pgfplotstableread[col sep=comma]{data/coverage-l32-median.csv}\covtable
\begin{figure}[t]
\centering
\begin{tikzpicture}
\begin{axis}[
  width=0.92\textwidth, height=7.4cm,
  xlabel={campaign time (hours)}, ylabel={distinct assertion violations (sum over 5 mazes)},
  xmin=0, xmax=8.2, ymin=0, ymax=180,
  xtick={0,1,...,8}, ytick={0,30,60,90,120,153},
  grid=major, grid style={gray!25},
  legend style={font=\small, cells={anchor=west}, at={(0.5,-0.2)}, anchor=north, legend columns=2},
  tick label style={font=\small}, label style={font=\small},
  const plot
]
\addplot[name path=wmax, draw=none, forget plot] table[x=hours,y=wake_max]{\covtable};
\addplot[name path=wmin, draw=none, forget plot] table[x=hours,y=wake_min]{\covtable};
\addplot[cwake!25, forget plot] fill between[of=wmax and wmin];
\addplot[name path=fmax, draw=none, forget plot] table[x=hours,y=foundry_max]{\covtable};
\addplot[name path=fmin, draw=none, forget plot] table[x=hours,y=foundry_min]{\covtable};
\addplot[cfoundry!20, forget plot] fill between[of=fmax and fmin];
\addplot[name path=emax, draw=none, forget plot] table[x=hours,y=echidna_max]{\covtable};
\addplot[name path=emin, draw=none, forget plot] table[x=hours,y=echidna_min]{\covtable};
\addplot[cechidna!20, forget plot] fill between[of=emax and emin];
\addplot[cwake, very thick] table[x=hours,y=wake]{\covtable};
\addlegendentry{Wake 5 (median, 5 seeds)}
\addplot[cfoundry, very thick] table[x=hours,y=foundry]{\covtable};
\addlegendentry{Foundry (median, 5 seeds)}
\addplot[cechidna, very thick] table[x=hours,y=echidna]{\covtable};
\addlegendentry{Echidna (median, 5 seeds)}
\addplot[cagent, very thick, dashed] coordinates {(0,0) (0.6167,0) (0.6167,150) (1.3167,150) (1.3167,153) (8.2,153)};
\addlegendentry{Agent-written harness, August campaign}
\addplot[cagent, only marks, mark=triangle*, mark size=2.2pt] coordinates {(0.55,153) (0.57,153) (0.65,153) (1.3167,153)};
\addlegendentry{Four further campaigns, completion}
\addplot[black!60, thin, dashed, forget plot] coordinates {(0,153) (8.2,153)};
\node[font=\scriptsize, anchor=south east] at (axis cs:8.1,158) {Benchmark maximum: 153 reachable assertions};
\end{axis}
\end{tikzpicture}
\caption{Distinct assertions violated on the five L32 mazes. Solid lines show medians over five seeds; shaded bands show the seed range. The dashed magenta step shows the August agent campaign; triangles mark completion of the four October campaigns. The benchmark maximum is 153 reachable assertions across all five mazes. Agent times include harness construction and solving.}
\label{fig:coverage}
\end{figure}
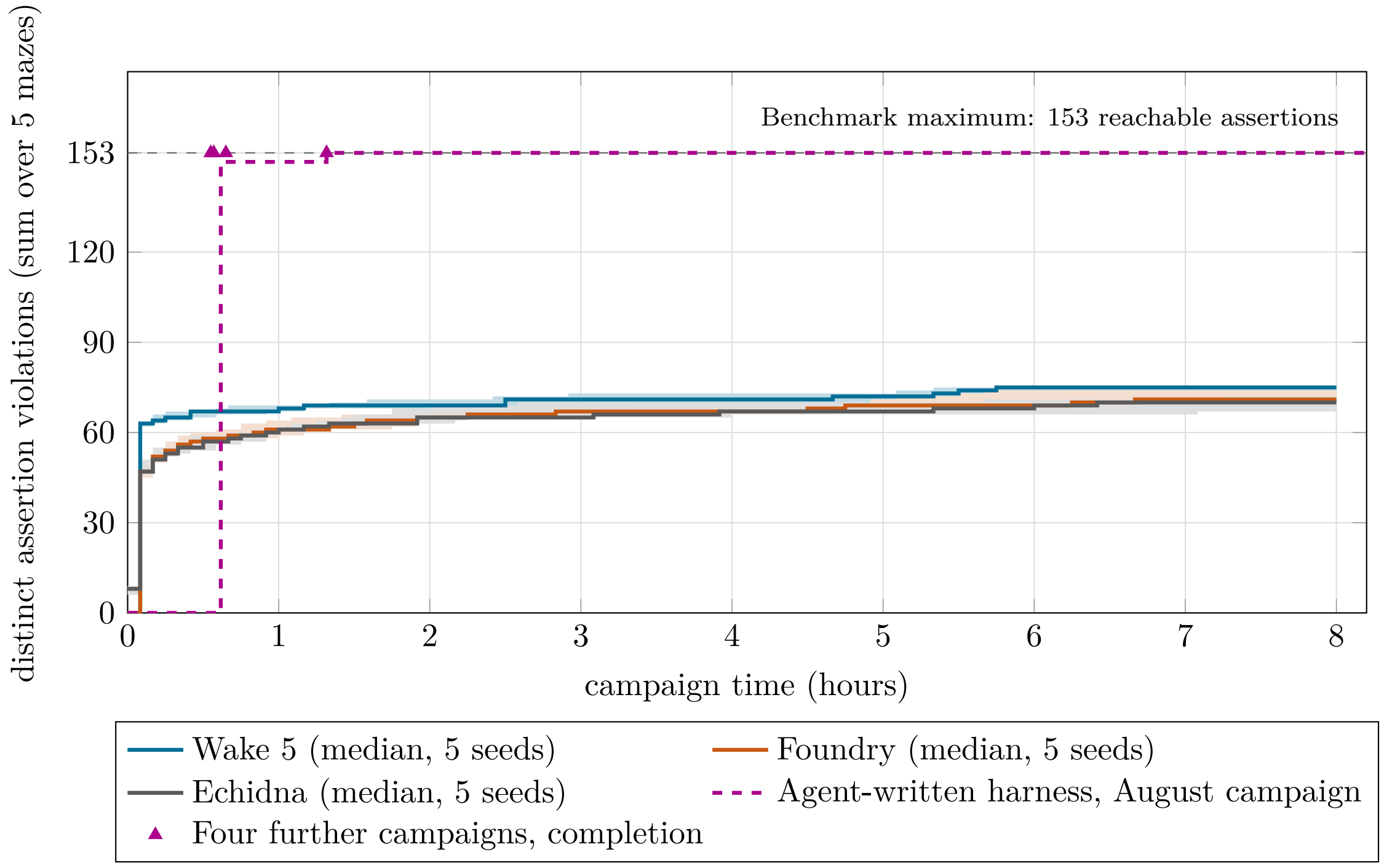

\begin{table}[t]
\centering\small
\caption{Assertion coverage across the five L32 mazes}
\label{tab:endpoints}
\setlength{\tabcolsep}{4pt}
\begin{tabular}{@{}l*{5}{r}r@{}}
\toprule
 & \multicolumn{5}{c}{Seed} & \\
\cmidrule(lr){2-6}
Tool & 0 & 1 & 2 & 3 & 4 & Median \\
\midrule
\wake{}~5 & 73 & 75 & 72 & 75 & 75 & \textbf{75} \\
Foundry & 74 & 72 & 71 & 70 & 71 & \textbf{71} \\
Echidna & 72 & 71 & 70 & 69 & 67 & \textbf{70} \\
\midrule
Agent-written harness & \multicolumn{5}{c}{$5\times$153, 33--79 min} & \textbf{153} \\
\bottomrule
\end{tabular}
\tablenote{Fuzzer entries are eight-hour totals per seed, out of 153 reachable assertions. All five agent campaigns reached 153 in 33--79 minutes.}
\end{table}

An 11.6$\times$ spread in throughput therefore bought at most five assertions of eight-hour coverage, and the two tools with equal throughput differ by four. Once the shallow assertions are exhausted, every tool waits on the same event: a generated input vector that satisfies a remaining guard chain. That probability is set by the generator, not the engine.

\subsection{RQ3: The generator prior}
\label{sec:prior}

Let $X$ be an unsigned 64-bit flow argument drawn by \wake{}'s generator (Section~\ref{sec:mgf}) and let $S=\{0,1,2^{64}-1,2^{63},2^{63}-1\}$. With $\kappa(x)=\max(1,\lceil\log_2(x+1)\rceil)$ the bit length of $x$,
\begin{align}
\Pr[X=x] \;&=\; 0.01\cdot\mathbf{1}[x\in S] \;+\; \frac{0.95}{64}\sum_{b=\kappa(x)}^{64} 2^{-b}, \label{eq:prior}\\
\Pr[X<2^{k}] \;&=\; 0.01\,\big|S\cap[0,2^k)\big| \;+\; \frac{0.95}{64}\Big(k+\sum_{b=k+1}^{64}2^{k-b}\Big). \label{eq:priorcdf}
\end{align}
Thirty-two of the 64 bit-length choices produce values below $2^{32}$, and longer draws can land there too, so $\Pr[X<2^{32}]=0.50984$: half of all draws fall in a range that uniform sampling reaches with probability $2^{-32}\approx 2.3\times10^{-10}$ (Figure~\ref{fig:prior}). For an exact 32-bit constant $c$, $\Pr[X=c]=6.9\times10^{-12}$ against $5.4\times10^{-20}$ under uniform sampling, a factor of $1.3\times10^{8}$. The prior is general \wake{} behaviour for type-hinted flow arguments, not benchmark code.

\pgfplotstableread[col sep=comma]{data/wake-prior-cdf.csv}\priortable
\begin{figure}[t]
\centering
\begin{tikzpicture}
\begin{axis}[
  width=0.8\textwidth, height=6.4cm,
  xlabel={$k$ (bits)}, ylabel={$\Pr[X<2^{k}]$},
  xmin=0, xmax=64, ymode=log, ymin=1e-20, ymax=2,
  xtick={0,8,...,64},
  grid=major, grid style={gray!25},
  legend pos=south east, legend style={font=\small, cells={anchor=west}},
  tick label style={font=\small}, label style={font=\small}
]
\addplot[cwake, very thick] table[x=k,y=wake]{\priortable};
\addlegendentry{\wake{}~5 generator, Eq.~\eqref{eq:priorcdf}}
\addplot[cechidna, very thick, dashed] table[x=k,y=uniform]{\priortable};
\addlegendentry{uniform \texttt{uint64}}
\draw[cagent, thick] (axis cs:32,1e-20) -- (axis cs:32,2);
\node[font=\scriptsize, anchor=south, rotate=90, cagent] at (axis cs:31,1e-9) {L32 constant bound};
\end{axis}
\end{tikzpicture}
\caption{Cumulative probability that an unsigned 64-bit value is below $2^k$, for \wake{} and uniform sampling. Changes in slope at $k=1$, $63$ and $64$ reflect the generator's edge values.}
\label{fig:prior}
\end{figure}
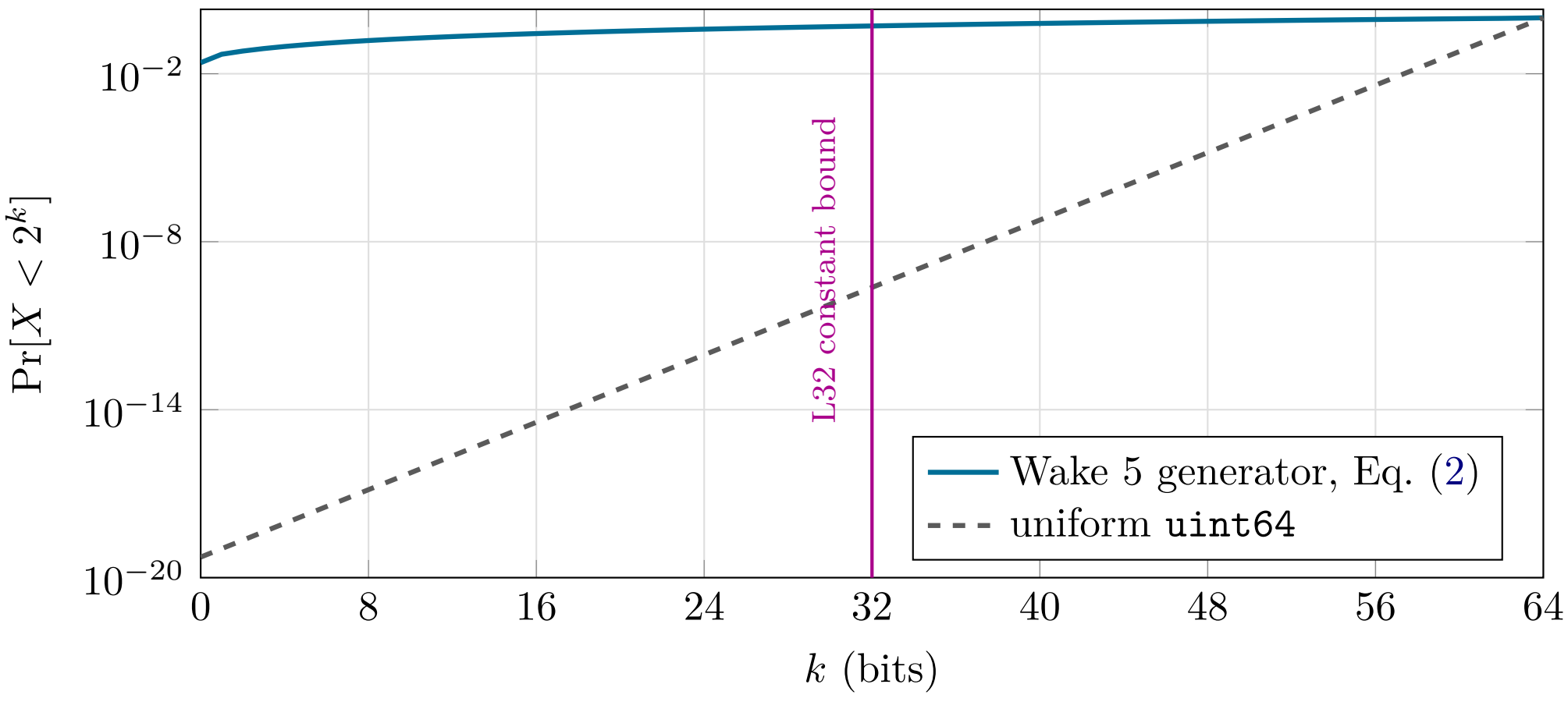

The prior explains why \wake{}'s unguided harness matches and, over five seeds, slightly exceeds Echidna's constant dictionary on L32, and it predicts where it will not: guards such as $p_0\neq 46$ are easier under uniform draws, and guards that require overflow ($p_4 > p_4\cdot p_6$) are never satisfied by small values. A supplementary eight-hour run on an L64 maze set with full-width constants, listed in Appendix~\ref{sec:supplementary}, is consistent with this prediction. The benchmark fork ships an experiment harness that isolates the prior by re-implementing it as an input arm alongside uniform, small-value and dictionary arms; those runs are ongoing and not reported here.

\subsection{RQ4: The agent-written harness}
\label{sec:agent}

A language-model agent was given the five L32 mazes and \wake{} and asked to find the assertions. The harness it produced over several develop--run--debug iterations is a Python \texttt{FuzzTest}, executed by \wake{}~5 like any other, whose input strategy is no longer random:

\begin{enumerate}
  \item \emph{Navigation from parsed source.} The agent parsed each maze's \texttt{step} function into a per-cell table of guard conditions, walls and assertion cells, and navigates by breadth-first search using transit inputs known to satisfy no assertion guard.
  \item \emph{Guard inputs from an SMT solver.} For each assertion cell the conjunction of guards is handed to Z3 with $p_0,\dots,p_7$ as 64-bit bit-vectors, preserving unchecked wrap-around. Queries that return \emph{unknown} are retried through a \texttt{simplify}, \texttt{bit-blast}, \texttt{sat} tactic, which rewrites the query to propositional satisfiability. Every satisfying model is independently re-evaluated in Python against the parsed conditions before it is retained.
  \item \emph{Replay as the oracle.} Each candidate is executed against a fresh deployment; only the replayed \texttt{Panic(0x01)}, the EVM panic code of a failed \texttt{assert}, counts. The solver result is a hypothesis, the execution is the evidence.
\end{enumerate}

\begin{table}[t]
\centering\footnotesize
\caption{Completion time and cost of agent-written harness campaigns}
\label{tab:agentcost}
\setlength{\tabcolsep}{3.5pt}
\begin{tabular}{@{}llrlrrrrr@{}}
\toprule
 & Date & 153 at & Ended & Wall & In, cached & In & Out & Cost \\
\midrule
1 & Aug 2026 & 79 min & on its own & 2 h 18 min & 39.09 & 0.46 & 0.28 & \$33.41 \\
2 & 5 Oct 2026 & 33 min & operator & 33 min & 5.40 & 0.15 & 0.10 & \$7.28 \\
3 & 5 Oct 2026 & 79 min & operator & 2 h 20 min & 25.47 & 0.80 & 0.22 & \$26.89 \\
4 & 5 Oct 2026 & 39 min & on its own & 3 h 29 min & 51.13 & 0.95 & 0.41 & \$45.94 \\
5 & 26 Oct 2026 & 34 min & on its own & 2 h 38 min & 44.63 & 0.56 & 0.33 & \$36.01 \\
\bottomrule
\end{tabular}
\tablenote{153 at: time to the last reachable assertion. Wall: total elapsed time, including work after the last assertion. Token counts are millions of Claude Opus tokens; In excludes cached input. Cost uses list prices and includes research-sub-agent usage where available; campaign~5 records failed researcher calls. Counts for operator-stopped campaigns end at the stop. Campaign 5 used a fixed protocol; its counts include developer and reviewer conversations, checked against their transcripts.}
\end{table}

The August campaign reached 150 of 153 assertions at minute 37 and all 153 at minute 79 (Figure~\ref{fig:coverage}). Its first 37 minutes were spent parsing and solving; the remaining assertions were those whose queries needed the bit-blasting fallback. Three further campaigns on 5~October 2026, on the same fixtures, also reached all 153 and completed at minutes 33, 39 and 79. Two of them emitted every finding within a second of each other once the harness ran, the solver having finished before execution started; the third repeated the August profile in two stages, 79 assertions by minute 34 and 146 by minute 56. A fifth campaign on 26~October 2026, run under a fixed protocol with its finding records, harness and token accounting retained, reached all 153 at minute 34: eight assertions at minute 30.6 in a first fuzz pass that aborted after eleven seconds on a harness error and sent the agent into debugging, and the remaining 145 at minute 34.2 in the third pass; 151 came from replaying solver-computed solutions and two from the harness's random walk. Across the five campaigns completion took 33 to 79 minutes (median 39), and the spread lies in harness construction and solving, not in execution. Completion is not the end of a campaign: with every assertion found, the agent kept proposing further work on a target that had nothing left to find. Three campaigns ended on their own 59, 123 and 170 minutes after their last assertion and two were stopped by the operator, so wall time ranged from 33 minutes to 3.5 hours and language-model cost from \$7 to \$46 (Table~\ref{tab:agentcost}), against a fixed eight hours per fuzzer campaign. A stopping rule is part of what the harness-construction workflow still lacks (Section~\ref{sec:discussion}). Random generation was replaced exactly where it is weakest, constructing inputs for arithmetic guards, and random walks were replaced by search over a parsed map; the runtime still deployed, transacted, checked and replayed at scale. The results are five campaigns of a different workflow, not a second engine measurement (Section~\ref{sec:threats}).

\subsection{RQ5: Static-analysis tools in an agentic review}
\label{sec:ablation}

To test whether deterministic code-structure facts help an agent, we ran an ablation in a different setting: a security review of an anonymised production proof-of-stake consensus client written in Go, with a Go-analysis MCP server that mirrors the surface of \wake{}'s server (workspace diagnostics, symbols, definitions, references, implementations, imports, direct callers and callees; eleven tools; no taint, reachability or vulnerability query)~\cite{prevratil2026ablation}. Three GPT-5.6 models (Sol, Terra, Luna) at the vendor's highest reasoning-effort setting (\texttt{xhigh}) each processed 168 cases, 128 questions naming a potential issue and 40 open-ended prompts, once with the tools available (\emph{tools on}) and once with the server, tool declarations and prompt references removed (\emph{tools off}). A supporting Luna instance at the \texttt{high} setting answered code-research sub-questions in every arm. The main comparison uses the 134 cases for which every model returned a usable verdict in both arms: 55 known bugs and 79 cases with no real bug; all 34 exclusions are non-bug cases. We report \emph{recall} (the share of the 55 known bugs found), \emph{correct rejection} (the share of the 79 non-bug cases dismissed, that is, false positives avoided), \emph{precision} (the share of reported issues that were real), and the cost of the language-model calls (\emph{LLM cost}).

\begin{table}[t]
\centering\small
\caption{Static-analysis tool ablation on 134 paired cases}
\label{tab:ablation}
\begin{tabular}{@{}llrrrrr@{}}
\toprule
Model & Metric & Tools on & Tools off & On $-$ off & $p$ & 95\% interval \\
\midrule
Sol   & recall            & 38/55 (69.1\%) & 39/55 (70.9\%) & $-1.8$\pp & 1.000 & $[-9.8,\,+6.1]$ \\
Terra & recall            & 32/55 (58.2\%) & 34/55 (61.8\%) & $-3.6$\pp & 0.727 & $[-13.7,\,+6.4]$ \\
Luna  & recall            & 32/55 (58.2\%) & 32/55 (58.2\%) & $0.0$\pp  & 1.000 & $[-8.7,\,+8.7]$ \\
\midrule
Sol   & correct rejection & 70/79 (88.6\%) & 69/79 (87.3\%) & $+1.3$\pp & 1.000 & $[-4.3,\,+6.8]$ \\
Terra & correct rejection & 70/79 (88.6\%) & 66/79 (83.5\%) & $+5.1$\pp & 0.289 & $[-1.9,\,+12.0]$ \\
Luna  & correct rejection & 67/79 (84.8\%) & 63/79 (79.7\%) & $+5.1$\pp & 0.219 & $[-0.9,\,+11.0]$ \\
\midrule
Sol   & precision         & 38/47 (80.9\%) & 39/49 (79.6\%) & $+1.3$\pp & -- & -- \\
Terra & precision         & 32/41 (78.0\%) & 34/47 (72.3\%) & $+5.7$\pp & -- & -- \\
Luna  & precision         & 32/44 (72.7\%) & 32/48 (66.7\%) & $+6.1$\pp & -- & -- \\
\bottomrule
\end{tabular}
\tablenote{Recall uses 55 known bugs; correct rejection uses 79 non-bug cases. Changes and normal-approximation 95\% intervals are for tools on minus tools off, in percentage points (pp). The $p$-values are from exact McNemar tests. Precision is descriptive: its denominator is the number of confirmed verdicts, which differs between arms.}
\end{table}

Recall did not improve for any model (Table~\ref{tab:ablation}); Sol and Terra found one and two more known bugs, respectively, with the tools off. Correct rejection of non-bug cases rose for all three, and with it precision, by 1.3 to 6.1 percentage points. No paired comparison reaches $p<0.05$; the smallest $p$ is 0.219 for Luna's rejections. Tool use varied widely (Table~\ref{tab:cost}): Luna called the tools in 96.4\% of runs, Terra in 32.1\%, so Terra's ablation is close to comparing an arm against itself. LLM cost rose with tool use, by 12.7\% for Luna, 8.1\% for Sol and 0.0\% for Terra, whose runs rarely called the tools; the static-analysis server ran locally and had no fee. A proof-of-concept arm (ten pre-registered bugs, the same three models, executable Go demonstrations) and an open-ended arm (a smaller open-weight model prompted only to perform a deep security analysis, ten runs per setting, blind-judged against 45 known bugs) were flat as well: 11/30 versus 12/30 failing demonstrations and 6 versus 5 bugs recovered, both $p=1.000$.

\begin{table}[t]
\centering\small
\caption{Static-analysis tool use and change in language-model cost}
\label{tab:cost}
\begin{tabular}{@{}lrrrr@{}}
\toprule
Model & Runs using tools & Calls per run & Median calls & LLM cost change \\
\midrule
Sol   & 128/168 (76.2\%) & 8.01  & 7.0 & $+8.1\%$ \\
Terra & 54/168 (32.1\%)  & 1.25  & 0.0 & $+0.0\%$ \\
Luna  & 162/168 (96.4\%) & 10.11 & 9.5 & $+12.7\%$ \\
\bottomrule
\end{tabular}
\tablenote{Tool-use statistics cover 168 tools-on runs per model. Calls / run is the mean; Median is the median number of calls. Cost changes compare tools on with tools off, averaged over the 134 paired cases in Table~\ref{tab:ablation}; they include supporting Luna calls. Prices verified 2026-08-04.}
\end{table}

The reading is that code-structure facts helped the models \emph{verify} (reject false positives) but not \emph{discover}. The tools had no query for the questions an audit turns on: whether untrusted input reaches a release path, which state changes precede an external call, whether an attacker-controlled implementation can satisfy an interface. A follow-up should test security-specific deterministic outputs, taint paths, attacker reachability, state-transition summaries and candidate invariants, against cases where each has a clear route to the answer.

\subsection{A datapoint from a real audit}
\label{sec:audit}

One anonymised engagement provides a non-synthetic reference for harnesses written by agents (Table~\ref{tab:audit}). The full human report contained 7 critical, 3 high and 4 medium findings. The manually guided fuzzing portion, 17 person-days of auditor time, accounted for 6 critical, 0 high and 1 medium of them. Two independent agent runs of the same fuzzing task, each writing and iterating a \wake{} harness, reported 4/1/1 in 84 elapsed hours and 5/3/2 in 28 elapsed hours. Effort and elapsed time are different units, so the rows are context; the controlled observation is the divergence between the two agent runs, where the run that took three times longer reported fewer findings in every severity.

\begin{table}[t]
\centering\small
\caption{Findings and time in one anonymised audit}
\label{tab:audit}
\begin{tabular}{@{}lrrrl@{}}
\toprule
Execution & Critical & High & Medium & Time \\
\midrule
Human report (all methods) & 7 & 3 & 4 & not recorded \\
Human MGF & 6 & 0 & 1 & 17 person-days \\
Agent run 1 & 4 & 1 & 1 & 84 h elapsed \\
Agent run 2 & 5 & 3 & 2 & 28 h elapsed \\
\bottomrule
\end{tabular}
\tablenote{Human MGF is the manually guided fuzzing subset of the full human report. Agent rows are independent runs of the same fuzzing task. Person-days measure human effort; elapsed hours measure agent runtime and are not directly comparable.}
\end{table}

\section{Discussion}
\label{sec:discussion}

\paragraph{Throughput is a prerequisite, not the constraint.}
\wake{}~5 closes the throughput gap between a Python-hosted framework and Rust-native fuzzers (RQ1), and the eight-hour curves show that the gap was not what separated tools on this benchmark (RQ2). On guard chains built from equalities against constants, every random generator waits on the same low-probability event, and the generator prior (RQ3) matters more than an order of magnitude of throughput. Speed still matters in two places: it shortens the develop--run--debug loop of harness construction, and it makes replay-based verification of thousands of solver candidates cheap. Both are on the path the agent took.

\paragraph{The fuzzer's role is moving from search to execution.}
The agent-written harness (RQ4) did not replace the fuzzer; it replaced the part of the workflow that decides what the fuzzer should try. That is the MGF division of labour of Section~\ref{sec:mgf} with the manual part automated, and it is why we consider a Python harness language and an in-process engine the right substrate for agents: the harness is ordinary code the agent can read, run and revise, and each revision executes at native speed.

\paragraph{What an agent-written harness inherits.}
Three failure modes appeared in the surrounding practice and are not measured by the benchmark. \emph{Correlated oracles}: a differential model derived from the implementation reproduces the implementation's defects. If a contract credits \texttt{msg.value * 2} by mistake and the agent translates that rule into the model, the equality assertion passes precisely where the contract is wrong; the oracle fails silently. An independent invariant, such as credited balances never exceeding assets held, exposes what translation cannot, and whether an agent reliably derives such invariants rather than restating source behaviour is open. \emph{Non-deterministic completion}: independent runs of the same task produce different harnesses with different differential-model depth and different assertions (Section~\ref{sec:audit}); a harness that builds, runs and passes carries no evidence that its security model is adequate, so a production workflow needs an explicit stopping rule and a quality bar separate from task completion. \emph{Priority inversion}: given a high-value, high-effort task (replace a mock with an integration test against forked live-chain state) and a low-value, self-contained one (implement a fee setter), agents were observed to start with the latter and defer the former behind a plausible rationale; priority adherence has to be evaluated separately from polished status reporting.

\paragraph{Deterministic helpers for non-deterministic reasoning.}
The ablation (RQ5) is a caution against inferring value from tool use: the models called the tools, the calls cost more, and recall did not move. Deterministic facts are a stable input to a variable reasoning process, not a guarantee of a better outcome. \wake{}'s MCP server is designed on that reading: it answers structural questions compactly and leaves the security question to whoever, or whatever, asks it.

\section{Threats to Validity}
\label{sec:threats}

\paragraph{Construct validity.}
Distinct assertion violations are Daedaluzz's own score; they measure reachability under a fuzzer, not bug severity or the cost of a miss. Transactions per second come from each tool's own counting and from active time (Section~\ref{sec:campaigns}), so throughput transfers as ratios only. In the ablation, precision is descriptive because its denominator differs by arm, and recall is measured against a fixed set of known bugs rather than against all bugs present.

\paragraph{Internal validity.}
Each tool ran its own idiomatic harness rather than a shared one, so harness shape is not fully separated from engine; the \wake{} harness was deliberately left unguided to keep the MGF method out of the measurement. The measurements used a release-candidate commit of \wake{}. The agent-written harness has five campaigns: completion ranged from 33 to 79 minutes, which shows the spread without estimating it, and three of the five record first-hit times during the fuzz run rather than replayed reproductions, with only their step logs retained, while the fifth retains its findings, harness and token accounting; two campaigns were stopped by the operator and the three that ended on their own did so long after completion, so the campaigns do not establish a stopping time, and the token counts of the stopped campaigns are as logged at the stop (Section~\ref{sec:campaigns}). The comparison is of a workflow, not an engine. Daedaluzz is public, so the agent's training data may include the benchmark family, although the L32 constants are freshly generated. The ablation has one run per case (four repeated cases changed one verdict).

\paragraph{External validity.}
Daedaluzz mazes are synthetic: unchecked arithmetic, capped constants and guard chains over eight parameters reward exactly the constraints an SMT solver dispatches, and no L32 campaign measured a real contract. Three fuzzers were run; Medusa, which Daedaluzz never wired in, and ItyFuzz, which led the Olympia comparison~\cite{chadt2024olympia} but is no longer actively developed, were not, and neither was any solver-assisted mode (Echidna's symbolic mode, Hybrid-Echidna, ItyFuzz's concolic executor), so the agent-written harness's advantage is not separated from that of a constraint solver as such. The generator-prior result (RQ3) is a statement about random input generation, not about fuzzers that solve constraints during the campaign. The ablation covers one Go codebase, one tool surface and three models from one vendor; it says nothing about security-specific static analysis or about \wake{}'s own MCP server on Solidity. The real-audit rows mix effort with elapsed time and cover one engagement.

\paragraph{Conclusion validity.}
Five seeds per tool are too few for narrow confidence intervals; we report the per-seed spread (67--75 across all tools), exact rank tests and effect sizes (Section~\ref{sec:coverage}), and treat a difference of a few assertions as small even where it is consistent across seeds. No paired comparison in the ablation reaches $p<0.05$, and with 55 known bugs the study is not powered to detect small recall effects; the conclusion that recall did not improve is a statement about this sample.

\section{Conclusion}
\label{sec:conclusion}

\wake{}~5 brings an in-process EVM to a Python-native testing framework and, on the Daedaluzz benchmark, matches the throughput of a Rust-native fuzzer. The same benchmark shows that on constant-guarded stateful contracts the binding constraint of random-input fuzzing has moved from throughput to input construction: an 11.6$\times$ throughput spread bought at most five assertions in eight hours, while a harness that derives inputs with a solver completed, in 33 to 79 minutes across five campaigns, a suite that eight-hour campaigns left more than half unfinished. A static-analysis tool ablation in an agentic review points the same way: code-structure facts helped agents verify, not discover. We read these results as a shift in the fuzzer's job description, from searching to executing what a human or an agent has decided to try, and we have designed \wake{}~5, its MGF harness model and its MCP server for that role. What remains open is a larger sample of agent runs under one measurement protocol, stopping rules for harness construction, and whether agents can derive oracles independent of the implementation under test.

\section*{Data and Software Availability}
\label{sec:availability}

\wake{} is available under the ISC licence at \mbox{\url{https://github.com/ack3-ai/wake}}. The measurements use commit \texttt{d33810e} (5.0.0rc2); the framework description covers commit \texttt{b46f5b5} (5.0.0rc3). Benchmark code is available at \url{https://github.com/ack3-ai/daedaluzz-benchmark}. Supplementary material provides the measured framework source, benchmark data, retained harnesses and analysis scripts. Underlying audit engagement records are withheld under client confidentiality; aggregate statistics are included. Finding records for three agent campaigns were not retained (Section~\ref{sec:method}).

\section*{Acknowledgements}

We thank Arman Ba\v{s}ovi\'{c} for the analysis and repair of the shrinking implementation on the \revm{} engine, ConsenSys Diligence for Daedaluzz, and the organisers of Web3 Security Summit Belgrade and ETHBelgrade 2026, where the measurements in Sections~\ref{sec:results} and~\ref{sec:ablation} were first presented.

\paragraph{AI assistance.}

Anthropic Claude Code and OpenAI Codex assisted the preparation and revision of manuscript text throughout this paper. Codex also assisted citation checking, LaTeX formatting, figure-caption revision, edits to the plotting code for Fig.~\ref{fig:coverage}, and preparation of the replication-package code and documentation. The agent-generated fuzzing harnesses are experimental objects described in Sections~\ref{sec:method} and~\ref{sec:agent}; they are distinct from assistance with preparing this manuscript. The authors remain responsible for the accuracy, originality and integrity of the paper.

\bibliographystyle{plainnat}
\bibliography{references}

\begin{thebibliography}{57}
\providecommand{\natexlab}[1]{#1}
\providecommand{\url}[1]{\texttt{#1}}
\expandafter\ifx\csname urlstyle\endcsname\relax
  \providecommand{\doi}[1]{doi: #1}\else
  \providecommand{\doi}{doi: \begingroup \urlstyle{rm}\Url}\fi

\bibitem[{ack3}(2026)]{wake2026repo}
{ack3}.
\newblock Wake: Python-based {Solidity} development, testing and fuzzing
  framework.
\newblock \url{https://github.com/ack3-ai/wake}, release candidate 5.0.0rc3,
  2026.
\newblock Documentation: \url{https://ack3.ai/wake/docs/5.0.0rc3/}.

\bibitem[{Ackee Blockchain}(2023{\natexlab{a}})]{ipor2023audit}
{Ackee Blockchain}.
\newblock {IPOR} protocol security audit report.
\newblock
  \url{https://github.com/Ackee-Blockchain/public-audit-reports/blob/master/2023/ackee-blockchain-ipor-protocol-report.pdf},
  2023{\natexlab{a}}.

\bibitem[{Ackee Blockchain}(2023{\natexlab{b}})]{pwn2023audit}
{Ackee Blockchain}.
\newblock {PWN} protocol v1.3 security audit report.
\newblock
  \url{https://github.com/PWNDAO/pwn_audits/blob/main/protocol/pwn-v1.3-ackee.pdf},
  2023{\natexlab{b}}.

\bibitem[{Ackee Blockchain}(2023{\natexlab{c}})]{pythonframeworksbenchmark}
{Ackee Blockchain}.
\newblock Python testing frameworks benchmark.
\newblock
  \url{https://github.com/Ackee-Blockchain/python-testing-frameworks-benchmark},
  2023{\natexlab{c}}.

\bibitem[{Ackee Blockchain}(2025)]{lido2025csm}
{Ackee Blockchain}.
\newblock Lido community staking module v2 audit report.
\newblock
  \url{https://github.com/lidofinance/audits/blob/main/Ackee%20Blockchain%20Community%20Staking%20Module%20v2%20Audit%20Report%2009-2025.pdf},
  2025.

\bibitem[Adams et~al.(2021)Adams, Zinsmeister, Salem, Keefer, and
  Robinson]{adams2021uniswapv3}
Hayden Adams, Noah Zinsmeister, Moody Salem, River Keefer, and Dan Robinson.
\newblock Uniswap v3 core.
\newblock Technical report, Uniswap, 2021.

\bibitem[{Alloy contributors}(2026)]{alloy}
{Alloy contributors}.
\newblock Alloy: High-performance {Ethereum} types and providers in {Rust}.
\newblock \url{https://github.com/alloy-rs/alloy}, 2026.

\bibitem[{Anthropic}(2024)]{mcp}
{Anthropic}.
\newblock Model context protocol.
\newblock \url{https://modelcontextprotocol.io/}, 2024.

\bibitem[{ApeWorX}(2026)]{ape}
{ApeWorX}.
\newblock Ape framework.
\newblock \url{https://apeworx.io/}, 2026.

\bibitem[Arcuri and Briand(2014)]{arcuri2014hitchhiker}
Andrea Arcuri and Lionel Briand.
\newblock A hitchhiker's guide to statistical tests for assessing randomized
  algorithms in software engineering.
\newblock \emph{Software Testing, Verification and Reliability}, 24\penalty0
  (3):\penalty0 219--250, 2014.
\newblock \doi{10.1002/stvr.1486}.

\bibitem[{bluealloy contributors}(2026)]{revm}
{bluealloy contributors}.
\newblock revm: Rust {Ethereum} virtual machine.
\newblock \url{https://github.com/bluealloy/revm}, 2026.
\newblock Version 41.0 used in this work.

\bibitem[B{\"o}hme et~al.(2016)B{\"o}hme, Pham, and
  Roychoudhury]{bohme2016aflfast}
Marcel B{\"o}hme, Van-Thuan Pham, and Abhik Roychoudhury.
\newblock Coverage-based greybox fuzzing as {Markov} chain.
\newblock In \emph{Proceedings of the 2016 ACM SIGSAC Conference on Computer
  and Communications Security (CCS)}, pages 1032--1043. ACM, 2016.
\newblock \doi{10.1145/2976749.2978428}.

\bibitem[Chadt et~al.(2024)Chadt, Hochrainer, W{\"u}stholz, and
  Christakis]{chadt2024olympia}
Jana Chadt, Christoph Hochrainer, Valentin W{\"u}stholz, and Maria Christakis.
\newblock Olympia: Fuzzer benchmarking for {Solidity}.
\newblock In \emph{Proceedings of the 39th IEEE/ACM International Conference on
  Automated Software Engineering (ASE)}, pages 2362--2365. ACM, 2024.
\newblock \doi{10.1145/3691620.3695352}.

\bibitem[Choi et~al.(2021)Choi, Kim, Kim, Grieco, Groce, and
  Cha]{choi2021smartian}
Jaeseung Choi, Doyeon Kim, Soomin Kim, Gustavo Grieco, Alex Groce, and Sang~Kil
  Cha.
\newblock Smartian: Enhancing smart contract fuzzing with static and dynamic
  data-flow analyses.
\newblock In \emph{Proceedings of the 36th IEEE/ACM International Conference on
  Automated Software Engineering (ASE)}, pages 227--239. IEEE, 2021.
\newblock \doi{10.1109/ASE51524.2021.9678888}.

\bibitem[Claessen and Hughes(2000)]{claessen2000quickcheck}
Koen Claessen and John Hughes.
\newblock {QuickCheck}: A lightweight tool for random testing of {Haskell}
  programs.
\newblock In \emph{Proceedings of the 5th ACM SIGPLAN International Conference
  on Functional Programming (ICFP)}, pages 268--279. ACM, 2000.
\newblock \doi{10.1145/351240.351266}.

\bibitem[{ConsenSys Diligence}(2023)]{daedaluzz2023}
{ConsenSys Diligence}.
\newblock Daedaluzz: Benchmark generator for smart-contract fuzzers.
\newblock \url{https://github.com/ConsenSysDiligence/daedaluzz}, 2023.
\newblock Accessed 2026-09-29.

\bibitem[de~Moura and Bj{\o}rner(2008)]{demoura2008z3}
Leonardo de~Moura and Nikolaj Bj{\o}rner.
\newblock {Z3}: An efficient {SMT} solver.
\newblock In \emph{Tools and Algorithms for the Construction and Analysis of
  Systems (TACAS)}, volume 4963 of \emph{LNCS}, pages 337--340. Springer, 2008.
\newblock \doi{10.1007/978-3-540-78800-3_24}.

\bibitem[Deng et~al.(2023)Deng, Xia, Peng, Yang, and Zhang]{deng2023titanfuzz}
Yinlin Deng, Chunqiu~Steven Xia, Haoran Peng, Chenyuan Yang, and Lingming
  Zhang.
\newblock Large language models are zero-shot fuzzers: Fuzzing deep-learning
  libraries via large language models.
\newblock In \emph{Proceedings of the 32nd ACM SIGSOFT International Symposium
  on Software Testing and Analysis (ISSTA)}, pages 423--435. ACM, 2023.
\newblock \doi{10.1145/3597926.3598067}.

\bibitem[{Echidna contributors}(2026)]{echidna230}
{Echidna contributors}.
\newblock Echidna 2.3.0 release notes (symbolic execution mode) and
  2.3.2-agents-preview-1 (mcp interface for fuzzing campaigns).
\newblock \url{https://github.com/crytic/echidna/releases/tag/v2.3.0};
  \url{https://github.com/crytic/echidna/releases/tag/v2.3.2-agents-preview-1},
  2026.
\newblock Released 2025-12-15 and 2026-01-20.

\bibitem[{eth-brownie contributors}(2023)]{brownie}
{eth-brownie contributors}.
\newblock Brownie: A {Python}-based development and testing framework for smart
  contracts.
\newblock \url{https://github.com/eth-brownie/brownie}, 2023.

\bibitem[Feist et~al.(2019)Feist, Grieco, and Groce]{feist2019slither}
Josselin Feist, Gustavo Grieco, and Alex Groce.
\newblock Slither: A static analysis framework for smart contracts.
\newblock In \emph{Proceedings of the 2nd International Workshop on Emerging
  Trends in Software Engineering for Blockchain (WETSEB)}, pages 8--15. IEEE,
  2019.
\newblock \doi{10.1109/WETSEB.2019.00008}.

\bibitem[{Foundry contributors}(2025)]{foundry130}
{Foundry contributors}.
\newblock Foundry v1.3.0 release notes: coverage-guided invariant fuzzing via
  \texttt{corpus\_dir}.
\newblock \url{https://github.com/foundry-rs/foundry/releases/tag/v1.3.0},
  2025.
\newblock Released 2025-07-31.

\bibitem[{Foundry contributors}(2026)]{foundry}
{Foundry contributors}.
\newblock Foundry: A blazing fast, portable and modular toolkit for {Ethereum}
  application development.
\newblock \url{https://github.com/foundry-rs/foundry}, 2026.
\newblock Version 1.7.1 used in this work.

\bibitem[Gattermayer(2026)]{gattermayer2026frontier}
Josef Gattermayer.
\newblock Benchmarking frontier models on unpublished audits.
\newblock ack3 research journal,
  \url{https://ack3.ai/research/benchmarking-frontier-models-on-unpublished-audits/},
  2026.
\newblock Published 2026-08-03.

\bibitem[Gattermayer et~al.(2023)Gattermayer, P{\v{r}}evr{\'a}til, and
  Kub{\v{e}}na]{gattermayer2023wake}
Josef Gattermayer, Michal P{\v{r}}evr{\'a}til, and Jan Kub{\v{e}}na.
\newblock Evaluating testing frameworks for {Ethereum} blockchain applications:
  Introducing {Wake}.
\newblock Technical report, Ackee Blockchain, 2023.
\newblock
  \url{https://ackee.xyz/Wake_testing_framework_Whitepaper_draft_1.pdf}.

\bibitem[Gattermayer et~al.(2026)Gattermayer, Kalivoda, and
  Ba{\v{s}}ovi{\'c}]{gattermayer2026incidents}
Josef Gattermayer, Jan Kalivoda, and Arman Ba{\v{s}}ovi{\'c}.
\newblock The ack3 {H1} 2026 {DeFi} incident dataset: Audit scope across 135
  security incidents.
\newblock arXiv preprint arXiv:2608.13792, 2026.

\bibitem[Grieco et~al.(2020)Grieco, Song, Cygan, Feist, and
  Groce]{grieco2020echidna}
Gustavo Grieco, Will Song, Artur Cygan, Josselin Feist, and Alex Groce.
\newblock Echidna: Effective, usable, and fast fuzzing for smart contracts.
\newblock In \emph{Proceedings of the 29th ACM SIGSOFT International Symposium
  on Software Testing and Analysis (ISSTA)}, pages 557--560. ACM, 2020.
\newblock \doi{10.1145/3395363.3404366}.

\bibitem[Jiang et~al.(2018)Jiang, Liu, and Chan]{jiang2018contractfuzzer}
Bo~Jiang, Ye~Liu, and W.~K. Chan.
\newblock {ContractFuzzer}: Fuzzing smart contracts for vulnerability
  detection.
\newblock In \emph{Proceedings of the 33rd ACM/IEEE International Conference on
  Automated Software Engineering (ASE)}, pages 259--269. ACM, 2018.
\newblock \doi{10.1145/3238147.3238177}.

\bibitem[Klees et~al.(2018)Klees, Ruef, Cooper, Wei, and
  Hicks]{klees2018evaluating}
George Klees, Andrew Ruef, Benji Cooper, Shiyi Wei, and Michael Hicks.
\newblock Evaluating fuzz testing.
\newblock In \emph{Proceedings of the 2018 ACM SIGSAC Conference on Computer
  and Communications Security (CCS)}, pages 2123--2138. ACM, 2018.
\newblock \doi{10.1145/3243734.3243804}.

\bibitem[Lee et~al.(2022)Lee, Kim, and Cha]{lee2022fuzzle}
Haeun Lee, Soomin Kim, and Sang~Kil Cha.
\newblock Fuzzle: Making a puzzle for fuzzers.
\newblock In \emph{Proceedings of the 37th IEEE/ACM International Conference on
  Automated Software Engineering (ASE)}. ACM, 2022.
\newblock \doi{10.1145/3551349.3556908}.

\bibitem[Li et~al.(2023)Li, {Ben Allal}, Zi, Muennighoff, Kocetkov, Mou,
  Marone, Akiki, Li, Chim, et~al.]{li2023starcoder}
Raymond Li, Loubna {Ben Allal}, Yangtian Zi, Niklas Muennighoff, Denis
  Kocetkov, Chenghao Mou, Marc Marone, Christopher Akiki, Jia Li, Jenny Chim,
  et~al.
\newblock {StarCoder}: May the source be with you!
\newblock \emph{Transactions on Machine Learning Research}, 2023.
\newblock arXiv:2305.06161.

\bibitem[Liu et~al.(2025)Liu, Xue, Wu, Sun, Li, Shi, and
  Liu]{liu2025propertygpt}
Ye~Liu, Yue Xue, Daoyuan Wu, Yuqiang Sun, Yi~Li, Miaolei Shi, and Yang Liu.
\newblock {PropertyGPT}: {LLM}-driven formal verification of smart contracts
  through retrieval-augmented property generation.
\newblock In \emph{Proceedings of the Network and Distributed System Security
  Symposium (NDSS)}. Internet Society, 2025.
\newblock \doi{10.14722/ndss.2025.241357}.

\bibitem[MacIver et~al.(2019)MacIver, Hatfield-Dodds, and {many other
  contributors}]{maciver2019hypothesis}
David~R. MacIver, Zac Hatfield-Dodds, and {many other contributors}.
\newblock Hypothesis: A new approach to property-based testing.
\newblock \emph{Journal of Open Source Software}, 4\penalty0 (43):\penalty0
  1891, 2019.
\newblock \doi{10.21105/joss.01891}.

\bibitem[McNemar(1947)]{mcnemar1947}
Quinn McNemar.
\newblock Note on the sampling error of the difference between correlated
  proportions or percentages.
\newblock \emph{Psychometrika}, 12\penalty0 (2):\penalty0 153--157, 1947.
\newblock \doi{10.1007/BF02295996}.

\bibitem[{Microsoft}(2024)]{lsp}
{Microsoft}.
\newblock Language server protocol specification.
\newblock \url{https://microsoft.github.io/language-server-protocol/}, 2024.

\bibitem[Nguyen et~al.(2020)Nguyen, Pham, Sun, Lin, and Minh]{nguyen2020sfuzz}
Tai~D. Nguyen, Long~H. Pham, Jun Sun, Yun Lin, and Quang~Tran Minh.
\newblock {sFuzz}: An efficient adaptive fuzzer for {Solidity} smart contracts.
\newblock In \emph{Proceedings of the 42nd International Conference on Software
  Engineering (ICSE)}, pages 778--788. ACM, 2020.
\newblock \doi{10.1145/3377811.3380334}.

\bibitem[{Nomic Foundation}(2026)]{hardhat}
{Nomic Foundation}.
\newblock Hardhat: {Ethereum} development environment for professionals.
\newblock \url{https://hardhat.org/}, 2026.

\bibitem[P{\v{r}}evr{\'a}til(2026{\natexlab{a}})]{prevratil2026ablation}
Michal P{\v{r}}evr{\'a}til.
\newblock Static analysis {MCP} tools in an {AI} audit benchmark.
\newblock ack3 research journal,
  \url{https://ack3.ai/research/ablating-static-analysis-from-an-ai-security-benchmark/},
  2026{\natexlab{a}}.
\newblock Presented at ETHBelgrade 2026 as ``Does static analysis actually help
  AI?''.

\bibitem[P{\v{r}}evr{\'a}til(2026{\natexlab{b}})]{prevratil2026fuzzers}
Michal P{\v{r}}evr{\'a}til.
\newblock Benchmarking {EVM} fuzzers against an agent-written harness.
\newblock ack3 research journal,
  \url{https://ack3.ai/research/benchmarking-evm-fuzzers-against-an-agent-written-harness/},
  2026{\natexlab{b}}.
\newblock Published 2026-08-26; presented at Web3 Security Summit Belgrade,
  2026-08-26.

\bibitem[{PyO3 contributors}(2026)]{pyo3}
{PyO3 contributors}.
\newblock {PyO3}: {Rust} bindings for the {Python} interpreter.
\newblock \url{https://github.com/PyO3/pyo3}, 2026.
\newblock Version 0.29 used in this work.

\bibitem[Regehr et~al.(2012)Regehr, Chen, Cuoq, Eide, Ellison, and
  Yang]{regehr2012creduce}
John Regehr, Yang Chen, Pascal Cuoq, Eric Eide, Chucky Ellison, and Xuejun
  Yang.
\newblock Test-case reduction for {C} compiler bugs.
\newblock In \emph{Proceedings of the 33rd ACM SIGPLAN Conference on
  Programming Language Design and Implementation (PLDI)}, pages 335--346. ACM,
  2012.
\newblock \doi{10.1145/2254064.2254104}.

\bibitem[Schloegel et~al.(2024)Schloegel, Bars, Schiller, Bernhard,
  Scharnowski, Crump, Ale-Ebrahim, Bissantz, Muench, and
  Holz]{schloegel2024sok}
Moritz Schloegel, Nils Bars, Nico Schiller, Lukas Bernhard, Tobias Scharnowski,
  Addison Crump, Arash Ale-Ebrahim, Nicolai Bissantz, Marius Muench, and
  Thorsten Holz.
\newblock {SoK}: Prudent evaluation practices for fuzzing.
\newblock In \emph{Proceedings of the 2024 IEEE Symposium on Security and
  Privacy (S\&P)}, pages 1974--1993. IEEE, 2024.
\newblock \doi{10.1109/SP54263.2024.00137}.

\bibitem[Shou et~al.(2023)Shou, Tan, and Sen]{shou2023ityfuzz}
Chaofan Shou, Shangyin Tan, and Koushik Sen.
\newblock Ityfuzz: Snapshot-based fuzzer for smart contract.
\newblock In \emph{Proceedings of the 32nd ACM SIGSOFT International Symposium
  on Software Testing and Analysis (ISSTA)}, pages 322--333. ACM, 2023.
\newblock \doi{10.1145/3597926.3598059}.

\bibitem[Shou et~al.(2024)Shou, Liu, Lu, and Sen]{shou2024llm4fuzz}
Chaofan Shou, Jing Liu, Doudou Lu, and Koushik Sen.
\newblock {LLM4Fuzz}: Guided fuzzing of smart contracts with large language
  models.
\newblock arXiv preprint arXiv:2401.11108, 2024.

\bibitem[{Solidity Team}(2026)]{solidity}
{Solidity Team}.
\newblock Solidity documentation.
\newblock \url{https://docs.soliditylang.org/}, 2026.

\bibitem[{Trail of Bits}(2025)]{slithermcp}
{Trail of Bits}.
\newblock slither-mcp: {Slither} analyses as {MCP} tools.
\newblock \url{https://github.com/trailofbits/slither-mcp}, 2025.

\bibitem[{Trail of Bits}(2026)]{medusa}
{Trail of Bits}.
\newblock Medusa: Parallelized, coverage-guided, mutational {Solidity} smart
  contract fuzzing.
\newblock \url{https://github.com/crytic/medusa}, 2026.

\bibitem[Vargha and Delaney(2000)]{vargha2000a12}
Andr{\'a}s Vargha and Harold~D. Delaney.
\newblock A critique and improvement of the {CL} common language effect size
  statistics of {McGraw} and {Wong}.
\newblock \emph{Journal of Educational and Behavioral Statistics}, 25\penalty0
  (2):\penalty0 101--132, 2000.
\newblock \doi{10.3102/10769986025002101}.

\bibitem[Wang et~al.(2026)Wang, Bigger, Xu, Lin, Applebaum, Patwardhan,
  Yukseloglu, and Watkins]{wang2026evmbench}
Justin Wang, Andreas Bigger, Xiaohai Xu, Justin~W. Lin, Andy Applebaum, Tejal
  Patwardhan, Alpin Yukseloglu, and Olivia Watkins.
\newblock {EVMbench}: Evaluating {AI} agents on smart contract security.
\newblock In \emph{Proceedings of the 43rd International Conference on Machine
  Learning}, volume 306 of \emph{Proceedings of Machine Learning Research},
  pages 125067--125099. PMLR, 2026.
\newblock URL \url{https://proceedings.mlr.press/v306/wang26k.html}.

\bibitem[Wang et~al.(2024)Wang, Pei, and Yang]{wang2024smartinv}
Sally~Junsong Wang, Kexin Pei, and Junfeng Yang.
\newblock {SmartInv}: Multimodal learning for smart contract invariant
  inference.
\newblock In \emph{Proceedings of the 2024 IEEE Symposium on Security and
  Privacy (S\&P)}, pages 2217--2235. IEEE, 2024.
\newblock \doi{10.1109/SP54263.2024.00126}.

\bibitem[Wood(2014)]{wood2014ethereum}
Gavin Wood.
\newblock Ethereum: A secure decentralised generalised transaction ledger.
\newblock Ethereum Project Yellow Paper, 2014.

\bibitem[Wu et~al.(2024)Wu, Li, Yan, Chen, Jiang, Wang, Luo, and
  Zhou]{wu2024arewethereyet}
Shuohan Wu, Zihao Li, Luyi Yan, Weimin Chen, Muhui Jiang, Chenxu Wang, Xiapu
  Luo, and Hao Zhou.
\newblock Are we there yet? {Unraveling} the state-of-the-art smart contract
  fuzzers.
\newblock In \emph{Proceedings of the 46th IEEE/ACM International Conference on
  Software Engineering (ICSE)}. ACM, 2024.
\newblock \doi{10.1145/3597503.3639152}.

\bibitem[W{\"u}stholz(2023)]{wustholz2023benchmarking}
Valentin W{\"u}stholz.
\newblock Benchmarking smart-contract fuzzers.
\newblock ConsenSys Diligence blog,
  \url{https://diligence.consensys.io/blog/2023/04/benchmarking-smart-contract-fuzzers/},
  2023.
\newblock Accessed 2026-09-29.

\bibitem[W{\"u}stholz and Christakis(2020)]{wustholz2020harvey}
Valentin W{\"u}stholz and Maria Christakis.
\newblock Harvey: A greybox fuzzer for smart contracts.
\newblock In \emph{Proceedings of the 28th ACM Joint European Software
  Engineering Conference and Symposium on the Foundations of Software
  Engineering (ESEC/FSE)}, pages 1398--1409. ACM, 2020.
\newblock \doi{10.1145/3368089.3417064}.

\bibitem[Xia et~al.(2024)Xia, Paltenghi, Tian, Pradel, and
  Zhang]{xia2024fuzz4all}
Chunqiu~Steven Xia, Matteo Paltenghi, Jia~Le Tian, Michael Pradel, and Lingming
  Zhang.
\newblock Fuzz4all: Universal fuzzing with large language models.
\newblock In \emph{Proceedings of the 46th IEEE/ACM International Conference on
  Software Engineering (ICSE)}. ACM, 2024.
\newblock \doi{10.1145/3597503.3639121}.

\bibitem[Zeller and Hildebrandt(2002)]{zeller2002simplifying}
Andreas Zeller and Ralf Hildebrandt.
\newblock Simplifying and isolating failure-inducing input.
\newblock \emph{IEEE Transactions on Software Engineering}, 28\penalty0
  (2):\penalty0 183--200, 2002.
\newblock \doi{10.1109/32.988498}.

\bibitem[Zhang et~al.(2024)Zhang, Zheng, Bai, Li, Ma, Xie, Li, Sun, and
  Liu]{zhang2024fuzzdriver}
Cen Zhang, Yaowen Zheng, Mingqiang Bai, Yeting Li, Wei Ma, Xiaofei Xie, Yuekang
  Li, Limin Sun, and Yang Liu.
\newblock How effective are they? {Exploring} large language model based fuzz
  driver generation.
\newblock In \emph{Proceedings of the 33rd ACM SIGSOFT International Symposium
  on Software Testing and Analysis (ISSTA)}, pages 1223--1235. ACM, 2024.
\newblock \doi{10.1145/3650212.3680355}.

\end{thebibliography}

\appendix
\section{Supplementary measurements}
\label{sec:supplementary}

\paragraph{In-process versus JSON-RPC.} An earlier internal run (July 2025) on the \emph{published} Daedaluzz mazes, twenty minutes and one CPU per campaign, summed over five mazes, gave Echidna~64, Foundry~58, \wake{} over Anvil via JSON-RPC~51, and \wake{} with the in-process engine~71 distinct assertions. Tool versions were not recorded and the run predates the history-bounding fixes; it indicates only that, within \wake{}, the engine change is worth the difference between last and first place in a short campaign.

\paragraph{L64 mazes.} Five eight-hour seeds on a maze set whose constants span the full 64-bit range gave per-seed sums (sorted) of Echidna 65/70/71/73/73 (median 71), Foundry 74/74/74/75/75 (median 74) and \wake{}~5 67/67/68/69/71 (median 68). The reachable ceiling of that set was not established with the solver, so the values are not normalised; the ordering is consistent with Section~\ref{sec:prior}: a small-value prior loses its advantage when constants are wide, and Foundry's bytecode-mined dictionary gains one.

\paragraph{Historical test-suite performance.} The 2023 measurement that motivated \wake{}'s architecture~\cite{gattermayer2023wake,pythonframeworksbenchmark} ran 271 Uniswap~v3~\cite{adams2021uniswapv3} test cases ported to each framework on a four-core cloud VM. Mean wall time on Anvil was 3.37~s for \wake{}, 10.96~s for Hardhat with Ethers.js, 34.80~s for Brownie and 53.27~s for Ape; on the Ganache chain the same suite took 15.78, 118.71, 51.48 and 72.72~s. Every framework ran over JSON-RPC in that measurement; the present work removes the socket from \wake{}'s default path.

\end{document}